\documentstyle[aps,preprint,axodraw]{revtex}
\def\btabl{\begin{table}}   \def\etabl{\end{table}}
\def\bea{\begin{eqnarray}}   \def\eea{\end{eqnarray}}
\def\bnn{\begin{eqnarray*}}   \def\enn{\end{eqnarray*}}
\def\beq{\begin{equation}}   \def\eeq{\end{equation}}  
\def\btabu{\begin{tabular}}   \def\etabu{\end{tabular}}
\def\bec{\begin{displaymath}} \def\eec{\end{displaymath}}
\def\nn{\nonumber}
\def\eqref#1{(\ref{#1})}

\renewcommand{\baselinestretch}{1.2}
\begin{document}
\newcommand{\bfig}{\begin{center}\begin{picture}}
\newcommand{\efig}[1]{\end{picture}\\{\small #1}\end{center}}
\newcommand{\flin}[2]{\ArrowLine(#1)(#2)}
\newcommand{\wlin}[2]{\DashLine(#1)(#2){3}}
\newcommand{\zlin}[2]{\DashLine(#1)(#2){5}}
\newcommand{\glin}[3]{\Photon(#1)(#2){2}{#3}}
\newcommand{\gsim}{\ \rlap{\raise 2pt\hbox{$>$}}{\lower 2pt \hbox{$\sim$}}\ }
\newcommand{\lin}[2]{\Line(#1)(#2)}
\newcommand{\sof}{\SetOffset}
\draft

%\date{\today}
\preprint{\vbox{\baselineskip=13pt
\rightline{CERN-TH/98-340}
\rightline{LPTHE Orsay-98/68}
\rightline{hep-ph/9812238}}}
\title{Low-energy photon--neutrino inelastic processes beyond the Standard
Model}
\author{A. Abada, 
J. Matias  and R. Pittau \footnote{
e-mail: abada@mail.cern.ch, matias@mail.cern.ch, pittau@mail.cern.ch.}} 
\vskip -0.5cm
\vspace{-0.5cm}
\address{{\small 
Theory Division, CERN, CH-1211 Geneva 23, Switzerland}}

\vskip -1cm
\maketitle \begin{abstract}  
We investigate in this work the leading contributions 
of  the MSSM with
$R$-parity  violation and of Left--Right models to the
low-energy 
five-leg photon--neutrino processes.
 We discuss the results and compare them to the
 Standard Model ones.
  
Pacs numbers: 13.10.+q, 13.15.+g, 14.70.Bh, 13.88.+e, 95.30.Cq.
\end{abstract}

\vspace{4.4cm}

\leftline{} 
\leftline{CERN-TH/98-340}
\leftline{December 1998}  

\renewcommand{\baselinestretch}{1.4}

\newpage \section{Introduction}

The low-energy photon--neutrino interactions are of  potential interest 
for astrophysics and cosmology (see
\cite{Chiu,Liu,Dic2,Dic3,ichep,nous,nous2}
and references therein). By low energies, we mean that the photons and the
neutrinos carry energies that  do not exceed the confinement scale. 
The  processes involving two neutrinos and two photons are strongly suppressed, 
not only because of the weak interaction but also because of Yang's theorem
 \cite{Yang}, which forbids a two photon-coupling to a  $J=1$ state.

The  processes involving one  more photon, such as 
\begin{eqnarray}
\gamma\nu    &\to& \gamma\gamma\nu     \label{eq4}\\
\nu\bar\nu   &\to& \gamma\gamma\gamma  \label{eq5}\\
\gamma\gamma &\to& \gamma\nu\bar\nu    \label{eq6}
\,, 
\end{eqnarray}
are not longer constrained by Yang's theorem. Moreover, 
the extra $\alpha$ in the five-leg 
cross section is compensated 
by the  replacement of the $\omega/M_W$ suppression  in the four-leg processes 
by an $\omega/m_e$ enhancement, $\omega$ being the centre-of-mass energy of the
collision.

Recently,  processes (\ref{eq4}), (\ref{eq5}) and  (\ref{eq6}) have
been studied,
first within an effective theory, see \cite{Dic3} and \cite{nous},
  based on the four-photon Euler--Heisenberg
Lagrangian \cite{EH}, which describes four-photon interactions.
 This effective approach gives reliable results for
energies below the threshold of $e^+e^-$ pair production,  while the
necessary
energy, interesting for the study of the supernova dynamics, is above $1$
MeV.
The extrapolation to energies  above $1$
MeV being suspect, the processes cited
above
have been computed directly in the Standard Model, see \cite{nous2} and
\cite{Dic1}.

 For astrophysical implications, 
 processes (\ref{eq4}) and (\ref{eq5}) may  give contributions to  the neutrino 
mean free path inside the supernova, and it is possible that process  
(\ref{eq6}) is an 
energy loss mechanism for stars. 

The conclusion  we arrived at in \cite{nous2} was that the five-leg
photon--neutrino
processes should be incorporated in supernova codes while they do not seem
to
have any  relevance in cosmology. Indeed, we have found that these
processes 
are unlikely to be
important for the study of the neutrino decoupling temperature
\cite{nous2}.

Knowing the contributions of these processes in the Standard Model, it is
worth
investigating whether they could have some importance and implications 
beyond the Standard Model.

Among possible minimal supersymmetric (MSSM) extensions of the Standard
Model 
(SM), the one including R-parity violating ($R\!\!\!/_p$) processes has
been
 attracting 
 increasing attention over the last years \cite{reviews}.
 In this paper, we compute the contribution of the process
$\gamma\nu\to\gamma\gamma\nu$ in the 
MSSM, with $R\!\!\!/_p$. 

Another popular extension of the SM  of the electroweak
 interactions
is based on the gauge group $SU(2)_{L} \times SU(2)_{R} \times U(1)_{\tilde
Y}$ \cite{all}, in which we  also compute the process
$\gamma\nu\to\gamma\gamma\nu$.

It is worth pointing out that the range of energy in which we will apply
the
above reaction is well below the mass of the ``exchanged particle", i.e. $W^\pm$ in
the case of
the Standard Model, $W^\pm$ or  $W'^\pm$ 
in the Left--Right model (LR),   and sleptons in the MSSM  with $R\!\!\!/_p$, 
so that the neutrino--electron coupling is treated as a
four-Fermi interaction. In all  cases, we perform the calculation with
massless neutrinos.

At low energies, far from the  confinement  scale $\sim 1$ GeV, 
the leading contribution
is given by diagrams involving only electrons or muons 
running inside the loop (see Fig. 1). 
 It is precisely the appearance of $m_{e}$ (or $m_\mu$) as a scale,
instead of $M_{W}$ (which is the scale governing the  
four-leg photon--neutrino reactions in the SM),
that makes such five-leg processes relevant at energies of the 
order of a few tens of MeV, in the SM.  In the MSSM, gauge invariance forbids the
 process of Fig. 1a with a
neutral scalar (sneutrino $\tilde\nu$) exchange; only Fig. 1b 
will thus contribute with  a slepton $\tilde l$ exchange, 
as we will see in section II. 
In the Left--Right model, the bosons $Z'$ and $W'^\pm$ enter the process 
in  Figs. 1a and 1b. 
 
The outline of the paper is as follows:   section II is devoted to the 
computation of  the five-leg
process in the MSSM with $R\!\!\!/_p$, while  the computation reported in section III
is
done in the LR model. Finally, we end up with a discussion on the comparison of these
  contributions with  the one obtained in the Standard Model.

\section{Computation of $\gamma\nu\to \gamma\gamma\nu$ in the MSSM with
$R\!\!\!/_p$} 

In SUSY extensions, gauge invariance and renormalizability no longer ensure
 lepton number $L$ (or baryon number $B$) conservation. 
 The generalization of the MSSM, which includes
 $R$-parity\footnote{$R_p$ is a multiplicative quantum number defined by 
 $R_p=(-1)^{L+3B+2S}$, where $L$, $B$, $S$ are the  lepton number,
 baryon number and spin of the particle, respectively.} 
 violation ($R\!\!\!/_p$),  allows the interaction 
  $\gamma\nu\to\gamma\gamma\nu$ to proceed via the
 exchange of a slepton,  unlike the   $W^\pm$ and $Z$
  boson exchange that takes place  in the Standard Model.

 The relevant $R\!\!\!/_p$ superpotential, consistent with
Lorentz invariance, gauge and SUSY symmetries \cite{reviews}, is,  
in what concerns our {\it low-energy}
process: 
\bea{\cal {W}}=\frac 1 2  \lambda_{ijk}L^iL^j\bar E^k\ ,\label{lag}\eea
where  $i$, $j$ and $k$ are  family indices and  $L^i$, $E^i$ are  the left-handed lepton
and
right-handed singlet charged  lepton superfield, respectively. 
The antisymmetry under $SU(2)_L$
 implies that the Yukawa
 couplings
are also antisymmetric under the exchange of the first two family indices, namely 
$\lambda_{ijk}=-\lambda_{jik}$ and as far as our analysis is concerned, $\lambda$ is assumed to be real.
All the following bounds on the couplings $\lambda_{ijk}$ are derived  within the  degenerate 
mass of sleptons and squarks \cite{Bounds} of $\tilde m\sim 100$ GeV: 
\bea
&&\lambda_{12k}<0.05\left({\tilde m\over 100\ {\mathrm{GeV}}}\right)\, \quad
{\mathrm{for}}\quad k=1,2,3\ ,\nn\\
&& \lambda_{13k}<0.06\left({\tilde m\over 100\ {\mathrm{GeV}}}\right)\, \quad
{\mathrm{for}}\quad  k=1,2\quad
{\mathrm{and}}\quad
 \lambda_{133}<0.004\left({\tilde m\over
 100\ {\mathrm{GeV}}}\right),\nn\\
&&\lambda_{23k}<0.06\left({\tilde m\over 100\ {\mathrm{GeV}}}\right)\,\quad
{\mathrm{for}}\quad  k=1,2,3\ \ .\label{lambda} \eea

Expressing the relevant part of the superpotential in (\ref{lag})
 in terms of component fields, we obtain the
Lagrangian:
\bea
{\cal {L}}_{LLE}=\lambda_{ijk}\bigl[\ \tilde \nu_L^i \ e_L^j\ \bar e_R^k \
 +\ \left(\tilde e_R^k\right)^c \ \nu_L^i \ e_L^j\ -\ 
  \tilde e_L^i\  \nu_L^j\ \bar e_R^k\ \bigr]\ +
{\mathrm{h. c}}.\label{lagexplicit}\eea
Notice that there is no term containing a neutrino $\nu$ and a sneutrino 
$\tilde\nu$ field simultaneously.
This implies that the process with a sneutrino exchanged in Fig. 1a is not allowed by
gauge invariance and that only  the last two terms in Eq.  (\ref{lagexplicit}) 
will contribute to the 
diagram of Fig. 1b of the five-leg 
low-energy photon--neutrino processes (\ref{eq4}), (\ref{eq5}) and (\ref{eq6}).

Concerning the second term in the Lagrangian (\ref{lagexplicit}), 
because of the antisymmetric nature of the coupling, the lepton-number 
violation is manifest:  
the family-type of the lepton running inside the loop is different from the one of the incoming
neutrino. 
For example, if  we assume that we are in a situation where 
$m_{{\tilde e}^k_L}\gg
m_{{\tilde e}^k_R}$, so that the third term of the Lagrangian (\ref{lagexplicit})
 is not relevant, then, when computing $\nu_e\gamma\to\nu_e\gamma\gamma$, 
 the running lepton is either a
 muon or a tau and, of
course, the muon contribution is the leading one.
In this case, the most relevant contribution may come from
the processes  where the running lepton is an electron: 
\bea\nu_\mu\gamma\to\nu_\mu\gamma\gamma \quad 
 {\mathrm{or}}\quad\nu_\tau\gamma\to\nu_\tau\gamma\gamma\ .\label{lnv}\eea
 However,  processes (\ref{lnv})  are not relevant for
 the study  of  supernova dynamics since in this case,   we are interested in
 the $\nu_e$-type.

 When we assume that $m_{{\tilde e}^k_L}\simeq
m_{{\tilde e}^k_R}$,  the third term in the Lagrangian (\ref{lagexplicit}) 
is always relevant
 since, for both processes (\ref{lnv}) and the process 
 $\nu_e\gamma\to\nu_e\gamma\gamma$, the running lepton
 inside the loop could be an electron. There is no constraint due to
 the antisymmetric nature of the coupling. It is the exchanged slepton that
  has to have  a  family-type different from the incoming neutrino. 
 
Moreover, since the lepton number is violated in this model,
  we can have also the
 following flavour-changing neutrino transitions:
\bea\nu_e\gamma\to\nu_\mu\gamma\gamma\ ,\ \ \ \nu_e\gamma\to\nu_\tau\gamma\gamma\quad 
 {\mathrm{and}}\quad \nu_\mu\gamma\to\nu_\tau\gamma\gamma\ .\label{lv}\eea
Processes  (\ref{lv})  could, in principle, 
  play some role  
 in the solar-neutrino puzzle, since the first two  processes in (\ref{lv}) correspond to 
 $\nu_e$-type suppression. 
 The numerical impact, though, 
 could be insignificant on account of low cross sections.

  Taking into account only the contributions where the running lepton inside
  the loop is an electron, we observe that it is straightforward  to adapt the existing tools of the
Standard Model computation  \cite{nous2}, as long as 
we express the $R_p$-violating effective four-fermion operators (keeping only  
the charged slepton $\tilde e$ exchange) in the same $(V-A)\times(V-A)$ 
form \cite{Gian}: 
   \bea
 L_{\mathrm{eff}}=&&{\lambda_{ijk}^2\over 2}\left[\left[
{1\over m_{\tilde e_R^k}^2}\left(\bar\nu_L^i\gamma^\mu \nu_L^i\right)\ 
 \left(\bar e_L^j\gamma^\mu e_L^j\right)
 - {1\over m_{\tilde e_L^i}^2}\left(\bar\nu_L^j\gamma^\mu \nu_L^j\right)\ 
 \left(\bar e_R^k\gamma^\mu e_R^k\right)-\right.\right.\nn\\
&&\left.\left.  {1\over m_{\tilde e_R^k}^2}\left(\bar e_L^i\gamma^\mu \nu_L^i\right)\ 
 \left(\bar \nu_L^j\gamma^\mu e_L^j\right)+\left[ i\leftrightarrow j\right]\right]
 \right]\ .
\label{4-Fermi}\eea
 The contribution of the $R\!\!\!/_p$ processes in Fig. 1b  thus  has the 
same $\left(V -A\right)\times\left(V -A\right)$ structure as the $W$
exchange in the Standard Model, so that the coupling $g$ is replaced by the Yukawa
coupling $\lambda_{ijk}$ and the $W$-propagator by the slepton's 
 ($M_W\to m_{\tilde
e^k}$) in the computation of the amplitude. For simplicity, we 
  will assume 
that  sleptons  are  degenerate in mass, $m_{\tilde e^1}= m_{\tilde e^2}
=m_{\tilde e^3} $. 

To have an indication on the contribution of the 
processes involving the slepton exchange,
 we are assuming from now on a degenerate mass for ${{\tilde e}_L}$ and 
 ${{\tilde e}_R}$.

In the following, we will make use of the notations of Ref. \cite{nous2},  
where we have computed in the Standard Model 
the amplitudes and cross sections for the processes (\ref{eq4}), (\ref{eq5}) and (\ref{eq6}),
 the $\nu$ being of any family-type. Here, 
 for the sake of comparison with the SM model results,  we will concentrate on 
  process 
(\ref{eq4}), that is, $\nu_i\gamma\to\nu_j\gamma\gamma$. 
We will give in the following the different amplitudes according to the
 family-type of the
neutrinos engaged in the process using the Lagrangian (\ref{lagexplicit}), 
in terms of the amplitudes computed in the Standard
 Model ($M^{\mathrm{SM}}$) with the appropriate changes (the couplings, the exchanged-boson and the running
 lepton inside the loop). More
precisely:\\
i) The coefficient\footnote{The coefficient $v_{e}$, which is directly
related to the $Z^0$-exchange,  is given by 
$v_{e}=-\frac 1 2+2 s_{W}^{2}$, 
where $s_W$ ($c_W$) is the sine (cosine)  of the Weinberg
angle.} $v_e$ corresponding to the $Z$ exchange (Fig. 1a) is set to zero
because, as already mentioned, only Fig. 1b will contribute  here. \\
ii) The mass $m_e$ in the SM amplitude is replaced by the one of the appropriate 
charged lepton running inside the loop.\\
iii) $1/\Delta_{W}\sim-1/M_W^2$ is replaced by $1/\Delta m_{\tilde
e}\sim -1/ m_{\tilde
e}^2$,
 $m_{\tilde e}$ being degenerate.
 
 We have then
\bea 
&&M(\nu_e\gamma\to\nu_e\gamma\gamma)= \sum_{k=1,3}\lambda_{k11}^2
{M^{\mathrm{SM}}
(\nu_e\gamma\to\nu_e\gamma\gamma)
\over g^2}\bigl[v_e\to 0,  M_W\to 
 m_{\tilde e}\bigr]
 \ ,\label{un}\\
&&M(\nu_\mu\gamma\to\nu_\mu\gamma\gamma)= \left(\sum_{k=1,3}\lambda_{21k}^2 
+\sum_{i=1,3}\lambda_{i21}^2\right) 
{M^{\mathrm{SM}}
(\nu_e\gamma\to\nu_e\gamma\gamma)
\over g^2}\bigl[v_e\to 0, M_W\to 
 m_{\tilde e}\bigr]
\ ,\label{deux}\\
&&M(\nu_\tau\gamma\to\nu_\tau\gamma\gamma)= \left(\sum_{k=1,3}\lambda_{31k}^2
+\sum_{i=1,3}\lambda_{i31}^2\right) 
{M^{\mathrm{SM}}
(\nu_e\gamma\to\nu_e\gamma\gamma)
\over g^2}\bigl[v_e\to 0,  M_W\to 
 m_{\tilde e}\bigr]
\ .\label{trois}
\eea

The transitions given in (\ref{lv}) are also
 computed as follows: 
\bea 
&&M(\nu_e\gamma\to\nu_\mu\gamma\gamma)
\ =\ \sum_{k=1,3}\lambda_{k11}
\lambda_{k21}
{M^{\mathrm{SM}}(\nu_e\gamma\to\nu_e\gamma\gamma)
\over g^2}\bigl[v_e\to 0,   M_W\to 
 m_{\tilde e}\bigr]\ ,\label{quatre}\\
&&M(\nu_e\gamma\to\nu_\tau\gamma\gamma)= \sum_{k=1,3}\lambda_{k11}
\lambda_{k31}
{M^{\mathrm{SM}}(\nu_e\gamma\to\nu_e\gamma\gamma)
\over g^2}\bigl[v_e\to 0,  M_W\to 
 m_{\tilde e}\bigr]
\ ,\label{cinq}\\
&&M(\nu_\mu\gamma\to\nu_\tau\gamma\gamma)= \!\left(\sum_{k=1,3}\lambda_{21k}
\lambda_{31k}+\!\sum_{k=1,3}\lambda_{k21}
\lambda_{k31}\!\right)\!
{M^{\mathrm{SM}}(\nu_e\gamma\to\nu_e\gamma\gamma)
\over g^2}\bigl[v_e\to 0,   M_W\to 
 m_{\tilde e}\bigr]\ .\label{six}
\eea

In Eqs. ({\ref{un}--{\ref{six}), we are neglecting the contributions where the running
lepton inside the loop is a muon or a tau.
When the running lepton is an
electron, in principle, since the computation is already done in the 
Standard Model
 and since the changes one needs to do in this case are global factors, 
 the computation in the MSSM with $R\!\!\!/_p$ is straightforward.

Using the definition $${g^2\over 8 M_W^2}= {{G_F}\over \sqrt 2}$$
and taking the following  couplings\footnote{The bounds on the couplings
$\lambda_{ijk}$ have been rescaled according to the new degenerate slepton mass 
limit given by ALEPH \cite{Limit}.} as degenerate   
 (see Eqs.  (\ref{lambda})):
\bea
&&\lambda_{12k}=\lambda_{123}\ \sim 0.04,\quad {\mathrm{for}} \quad k=1,2,3\
,\nonumber\\
&&\lambda_{131}=\lambda_{132}\ \sim 0.05,\quad {\mathrm{and}}\ \
\lambda_{133}\ \sim 0\ ,\nonumber\\
&&\lambda_{23k}=\lambda_{233}\ \sim 0.05,\quad {\mathrm{for}} \quad
k=1,2,3\ ,\label{rescale}
\eea
 we find the following ratios of the cross 
section
 in the MSSM with $R\!\!\!/_p$ versus the SM ones:

\bea 
&&{\sigma^{\mathrm{MSSM}}(\nu_e\gamma\to\nu_e\gamma\gamma)\over
\sigma^{\mathrm{SM}}(\nu_e\gamma\to\nu_e\gamma\gamma)}
\simeq 1+{\left({\displaystyle{\sum_{k=1,3}}}\lambda_{k11}^2\right)^2\over  g^4 } {1\over (1+v_e)^2} 
{M_W^4\over m_{\tilde e}^4}  \ \sim 1+ 0.01\%
\ ,\label{crossun}\\
&&{\sigma^{\mathrm{MSSM}}(\nu_\mu\gamma\to\nu_\mu\gamma\gamma)\over
\sigma^{\mathrm{SM}}(\nu_\mu\gamma\to\nu_\mu\gamma\gamma)}
\simeq  1+{ \left({\displaystyle{\sum_{k=1,3}}}\lambda_{21k}^2 
+\displaystyle{\sum_{i=1,3}}\lambda_{i21}^2\right)^2 \over  g^4 } {1\over v_e^2} 
{M_W^4\over m_{\tilde e}^4}  \ \sim 1+32\%
\ ,\label{crossdeux}\\  
&&{\sigma^{\mathrm{MSSM}}(\nu_\tau\gamma\to\nu_\tau\gamma\gamma)\over
\sigma^{\mathrm{SM}}(\nu_\tau\gamma\to\nu_\tau\gamma\gamma)}
\simeq 1+{\left({\displaystyle{\sum^{}_{k=1,3}}}\lambda_{31k}^2
+{\displaystyle{\sum_{i=1,3}}}\lambda_{i31}^2\right)^2 \over  g^4 } {1\over
v_e^2} 
{M_W^4\over m_{\tilde e}^4}   \ \sim 1+ 40\%
\ .\label{crosstrois}
\eea 
When the final neutrino family-type
 is different from the initial one, we get:
 \bea 
&&{\sigma^{\mathrm{MSSM}}(\nu_e\gamma\to\nu_\mu\gamma\gamma)\over
\sigma^{\mathrm{SM}}(\nu_e\gamma\to\nu_e\gamma\gamma)}\simeq1+{
\left(\displaystyle{\sum^{}_{k=1,3}}\lambda_{k11}
\lambda_{k21}\right)^2\over  g^4 } {1\over
(1+v_e)^2} 
{M_W^4\over m_{\tilde e}^4}
\sim 1+0.004\%
\ ,\label{crossquatre}\\
&&{\sigma^{\mathrm{MSSM}}(\nu_e\gamma\to\nu_\tau\gamma\gamma)\over
\sigma^{\mathrm{SM}}(\nu_e\gamma\to\nu_e\gamma\gamma)}
\simeq 1+ {\left(\displaystyle{\sum^{}}_{k=1,3}\lambda_{k11}
\lambda_{k31}\right)^2\over  g^4 } {1\over
(1+v_e)^2} 
{M_W^4\over m_{\tilde e}^4}  \sim 1+0.003\%
\ ,\label{crosscinq}\\
&&{\sigma^{\mathrm{MSSM}}(\nu_\mu\gamma\to\nu_\tau\gamma\gamma)\over
\sigma^{\mathrm{SM}}(\nu_\mu\gamma\to\nu_\mu\gamma\gamma)}
\simeq 1+{\left(\displaystyle{\sum_{k=1,3}}\lambda_{21k}
\lambda_{31k}+\displaystyle{\sum_{k=1,3}}\lambda_{k21}
\lambda_{k31}\right)^2\over  g^4 } {1\over
v_e^2} 
{M_W^4\over m_{\tilde e}^4}    \sim 1+ 0.03\%
\ .\label{crosssix}
\eea 

The limits on the slepton masses are taken from the latest   ALEPH analysis
\cite{Limit}, which gives $m_{\tilde e}\sim 80$ GeV for the degenerate
 mass of the  sleptons; the bounds on the 
couplings $\lambda_{ijk}$ are given by Eq. (\ref{rescale}) \cite{Bounds}.
The Standard Model cross sections are computed in Refs. \cite{nous,nous2}.

One can see from Eqs. (\ref{crossquatre}), (\ref{crosscinq}) and (\ref{crosssix})
that  an electron-neutrino transforms
 better in a muon-neutrino than in a tau-neutrino. However, in view 
 of the smallness of the MSSM with $R\!\!\!/_p$ cross sections, these
 transitions are unlikely to be relevant to the solar neutrino puzzle, although
  the neutrino energy is in
  the appropriate range
    (from $0.1$ to $\sim 15$ MeV). On the other hand, in  these processes, 
  a muon-neutrino prefers to convert into 
 a tau-neutrino 
  rather than an electron-neutrino. 
  This goes in the direction of the Superkamiokande result; however, our processes
  hold for energies much below the energy of the atmospheric neutrinos ($>1$ GeV).

 We have made the same computation in the case where $m_{{\tilde e}^k_L}\gg
m_{{\tilde e}^k_R}$, so that the second term
of the  Lagrangian (\ref{lagexplicit})
 is the most  relevant and we have found that the  $R\!\!\!\!/_p$ MSSM cross
 sections 
 are enhanced by a factor of the order of $10\%$ at best, 
 relative to the SM cross sections.

It is worth emphasizing that the same combinations of the couplings 
 $\lambda_{ijk}$ that contribute to flavour-changing neutrino transitions (\ref{lv}) might  contribute as well to
flavour-changing charged-lepton radiative transitions such as
$$\mu\to e \gamma\ ,\quad \tau\to e\gamma\ \quad\mathrm{and}\quad
\tau\to\mu\gamma\ ,$$
 decays that are very restricted by experiment \cite{ejpc}. In this model,
a rough estimate of  the cross section of, say,  the $\mu\to e \gamma$
 transition, which occurs 
through the sneutrino exchange, gives $\sim 3\times 10^{-13}$ while the
 experimental bound is $\sim
10^{-11}$ \cite{ejpc}. 

Finally, the natural question that has to be addressed is how large  the
contribution of R-parity-conserving SUSY to the transitions
$\nu\gamma\to\nu\gamma\gamma$ could be. The answer is that the cross sections
then are 
smaller, since the particles in the loop are  heavier, as  is the case when the
running fermion inside the loop is a chargino.

\newpage\section{Computation of $\gamma\nu\to \gamma\gamma\nu$ in the 
Left--Right Model}
Left--Right models 
 are  based on the gauge group $SU(2)_{L} \times SU(2)_{R} \times
U(1)_{\tilde
Y}$ \cite{all,dfmz}. This is a natural framework to 
embed extra $W^\pm$ and $Z$ gauge 
bosons that could be  found in forthcoming colliders.
 For a more general case
with more than one generation of $W'^\pm$ and $Z'$,  see
Ref. \cite{qv}.
We will keep the notation used in \cite{dfmz}.

The leading contribution of this type of models to our processes consists 
mainly, 
of the substitution of the $W^\pm$ and $Z$ propagators by the corresponding
$W'^\pm$ and $Z'$, and the modification of the standard 
couplings of
 $W^\pm$ and 
$Z$ gauge bosons  with fermions because of  the mixing effect.

The charged-current gauge interactions of leptons are given by
\bea
\label{chgau}
{\cal L}_{CC} = \left( \begin{array}{cc} J_L^{\mu +} & J_R^{\mu +}
\end{array} \right) \left( \begin{array}{c}  W_{\mu  \, L}^-
\\ W_{\mu \, R}^- \end{array} \right) + {\rm h.c.}
= \left( \begin{array}{cc} J_W^{\mu +} & J_{W'}^{\mu  +}
\end{array} \right) \left( \begin{array}{c}  W_{\mu}^-
\\ {W'_{\mu}}^- \end{array} \right) + {\rm h.c.} \, ,
\eea
where
\bea
\label{ccurr}
J_W^{\mu +} = \cos \alpha_{\pm} J_L^{\mu +} + \sin \alpha_{\pm} J_R^{\mu
+} \, ,
\;\;\;
J_{W'}^{\mu +} = - \sin \alpha_{\pm} J_L^{\mu +} + \cos \alpha_{\pm}
J_R^{\mu +} \, ,
\eea
and the charged current associated with 
$SU(2)_L$ and $SU(2)_R$ lepton interactions are, respectively:
\bea
\label{chcur}
J^{\mu +}_L  = {g_L \over \sqrt{2}} \left(
\overline{e_L} \gamma^{\mu} \nu_L  \right) \, \, \, {\rm and} \, \, \,
J^{\mu +}_R  = {g_R \over \sqrt{2}} \left(
\overline{e_R} \gamma^{\mu} \nu_R  \right)\ .
\eea
From now on, 
we will call  A and B the Standard Model amplitudes corresponding to\\ 
 Figs. 1a  and 1b, respectively, which are given in \cite{nous}.

~From Eqs. (\ref{ccurr}) and (\ref{chcur}), and using the definition of the
mixing angle of Ref. \cite{dfmz}, we find that type B diagrams  (Fig. 1b),
now called $B^{W}$, 
 are modified through the mixing by:
\bea \label{bw}
B^{W}=\cos^2 \alpha_{\pm} B. 
\eea
Also a new diagram (called $B^{W'}$), with a $W'^\pm$ 
instead of the $W^\pm$ boson, appears:
\bea \label{bwp}
B^{W'}=\sin^2 \alpha_{\pm} {M_W^2 \over M_{W'}^2} B \ .
\eea

Similarly, the modification in the neutral-current gauge interactions (see
\cite{dfmz}) gives rise to a modification of type A diagrams (Fig. 1a): 
\bea
A^{Z}=\cos^2 \alpha_{0} A 
\eea
and to a  new contribution, due to the $Z'$, given by:
\bea
A^{Z'}=\sin^2 \alpha_{0} {M_Z^2 \over M_{Z'}^2} A \ .
\eea

The total contribution of the equivalent set of diagrams of Eq. (15)
in \cite{nous} or Eq. (10) in \cite{nous2} is then
\bea \label{eqt}
A_{123}^{Z}+A_{321}^{Z}+B_{123}^{W}+B_{321}^{W}+
A_{123}^{Z'}+A_{321}^{Z'}+B_{123}^{W'}+
B_{321}^{W'}=C_{LR}\Gamma_{\mu} L_{1}\ ,
\eea
where
\bea \label{clr}
C_{LR}=
{g_{L}^5 s_{W}^{3} \over 2} \left[\left(
\cos^2 \alpha_{\pm} + \sin^2 \alpha_{\pm} {M_{W}^{2}  \over
M_{W'}^2}
\right) + \rho \ v_{e}\left(
\cos^2 \alpha_{0} + \sin^2 \alpha_{0} {M_{Z}^{2}  \over M_{Z'}^2}
\right) \right] {1 \over M_{W}^2} \ ,
\eea
where $\rho={M_{W}^{2} / c_{W}^2 M_{Z}^2}$;  $L_{1}$ and $\Gamma_{\mu}$
are given in \cite{nous}; $L_{1}$  can be evaluated
in
the large $m_{e}$ limit, as in \cite{nous}, or exactly as in \cite{nous2}.
 The  corresponding SM coefficient $C_{SM}$ can be
obtained trivially from Eq.(\ref{clr}) in the limit $\alpha_{\pm}
\rightarrow 0$
and $\alpha_{0} \rightarrow 0$.

Notice that the New Physics contribution  enters, according to Eq. (\ref{eqt}), as 
a multiplicative factor. 

In order to evaluate $C_{LR}$, we will work in the
small mixing-angle-approximation 
substituting $\cos \alpha_{\pm,0} \rightarrow 1$
and
$\sin \alpha_{\pm,0}\rightarrow \alpha_{\pm,0}$. 
At this order, the  $\rho$ parameter that enters $C_{LR}$ is given
by \cite{dfmz}
\bea \label{rho}
\rho=1- \alpha_{\pm}^2 \left({M_{W'}^2 - M_{W}^2 \over M_{W}^2}
\right)
+\alpha_{0}^2 \left({M_{Z'}^2 - M_{Z}^2 \over M_{Z}^2} \right)\ .
\eea
Finally,  we should define our input parameters. Concerning   the sector 
of the model
that affects our computation,  we have only  eight parameters, which we will
choose to be $\alpha$, $M_{W}$ (or $G_F$), $M_{Z}$ and $m_{e}$ (as in the
Standard Model) plus $x=g_{R}/g_{L}$, $M_{W'}$, $\alpha_{\pm}$ and
$\alpha_{0}$. In terms of these, in the small-mixing-angle approximation, 
we find, to the precision required for the evaluation of Eq. (\ref{clr}), 
that:
\bea
s_{W}^{2}=1-{M_{W}^{2} \over M_{Z}^{2}} - {M_{W}^{2} \over M_{Z}^{2}}
\left(
\alpha_{\pm}^2 \left({M_{W'}^2 - M_{W}^2 \over M_{W}^2} \right)
-\alpha_{0}^2 \left({M_{Z'}^2 - M_{Z}^2 \over M_{Z}^2} \right)
\right)
\eea
and
\bea
M_{Z'}^2={x^{2} M_{W}^2 M_{W'}^2 -
{(M_{Z}^{2}-M_{W}^{2})}^{2}
\over x^{2} M_{W}^2 -M_{Z}^2 + M_{W}^2}
\eea
and, of course, $g_L=\sqrt {4 \pi \alpha} / s_{W}$.
The stronger bounds \cite{ejpc} on our input parameters comes  from the flavour-changing
neutral-current FCNC (mainly the 
$K_{L}-K_{S}$ mass difference)
\cite{soni,lang}, 
but they depend on the assumptions of the model (manifestly LR
symmetric models
\cite{mlr}, $g_{L} \neq g_{R}$ models \cite{lang}, fermiophobic models \cite{dfmz,qv},
etc.). For instance, 
in manifestly symmetric Left-Right models with $g_{L}=g_{R}$, the bound
on  the mass of the $W'$ is $M_{W'}\gsim 1.6$ TeV. The bounds on the mixing angles
depend on the CP-violating phases of the theory; for small phases, it is
$|\alpha_{\pm}|<0.0025$;  for large phases, it is $|\alpha_{\pm}|<0.033$.
If  the constraint $g_{L}=g_{R}$ is  relaxed, the bounds on  $M_{W'}$ masses
are much  weaker. Finally, a fermiophobic model, which automatically
guarantees the absence of FCNCs at tree level, allows for a relatively light 
$M_{W'}$ with no contradiction
with experimental data.

For a typical set of values of the second half of input parameters 
($x=g_{R}/g_{L}$, $M_{W'}$, $\alpha_{\pm}$ and
$\alpha_{0}$), taking 
into account the bounds on FCNCs for each model and the stringent bounds
on $\rho$ of Eq.(\ref{rho}), it is possible to obtain correction of at most
 a few per mille  
 to the SM value of $C_{SM}$, which means also a few per mille enhancement in the
 cross sections.

\section{Conclusions}

The cross sections computed in the $R\!\!\!\!/_p$ MSSM
 are enhanced by a factor of the order of few $10\%$ at best 
 relative to the SM cross sections,  while  the correction 
  in LR models is negligible. 

Concerning the cosmological implications, in view of the small 
enhancements found, 
 the conclusion that these low-energy five-leg processes
 are  not relevant to the study of the
 neutrino  decoupling temperature \cite{nous2} remains unchanged. 
 
The results found in these extensions of the SM  will also enforce 
 the conclusions made in \cite{nous2} on the supernova question, that is: 
these low-energy processes should be taken into account
 in the supernova codes.  

On the other hand, if the family-type of the neutrino changes in
 the transition,
 as it is allowed in the MSSM with $R\!\!\!/_p$, 
  the results found in this model then go in the direction of the
 actual conjecture, explaining results from both solar and atmospheric 
 neutrino experiments, see Eqs. 
 (\ref{crossquatre}), (\ref{crosscinq}) and (\ref{crosssix}). 
 However, the cross sections are far too small to have any substantial effect
 on the neutrino fluxes.

\section*{Acknowledgements}
We are specially indebted to G. Bhattacharyya for reading the
manuscript and making important comments. 
We thank  S. Davidson, E. Dudas, and  G. F. Giudice 
 for helpful
discussions. We warmly thank S. Lola for reading the paper.\\
J.M. acknowledges the financial support from a
Marie Curie EC Grant (TMR-ERBFMBICT 972147).

\vskip 4cm

% This is figure n. 2:
%----------------------------------------------------------------
\noindent \bfig(300,120)
\sof(-50,20)
\flin{40,0}{110,0}  \flin{110,0}{180,0}
\wlin{110,0}{110,20}\
\flin{110,20}{90,40}
\flin{90,40}{110,60}
\flin{110,60}{130,40}
\flin{130,40}{110,20}
\glin{90,40}{65,40}{3}
\glin{110,60}{110,85}{3}
\glin{130,40}{155,40}{3}
\Text(63,-5)[t]{$~~~~\nu$}
\Text(155,-5)[t]{$\nu~~~~$}
\Text(115,12)[l]{$Z,Z'$}
\Text(98,55)[r]{$\mu,e$}
\Text(115,70)[l]{$\gamma$}
\Text(77,36)[t]{$\gamma$}
\Text(143,36)[t]{$\gamma$}
\Text(60,40)[r]{$~\epsilon_1$}
\Text(160,40)[l]{$~\epsilon_3$}
\Text(110,90)[b]{$~\epsilon_2$}
\Text(150,87)[bl]{$(a)$}
\sof(140,20)
\flin{40,0}{80,0}  \wlin{80,0}{140,0} \flin{140,0}{180,0}
\flin{80,0}{90,20} \flin{90,20}{110,32}
\flin{110,32}{130,20} \flin{130,20}{140,0}
\glin{90,20}{67,37}{3}
\glin{110,32}{110,62}{3}
\glin{130,20}{153,37}{3}
\Text(63,-5)[t]{$~~~~\nu$}
\Text(155,-5)[t]{$\nu~~~~$}
\Text(110,4)[b]{$W,W',\tilde e$}
\Text(98,33)[r]{$\mu,e$} 
\Text(115,50)[l]{$\gamma$}
\Text(75,23)[t]{$\gamma$} 
\Text(145,23)[t]{$\gamma$}
\Text(65,45)[r]{$~\epsilon_1$}
\Text(155,45)[l]{$~\epsilon_3$}
\Text(110,70)[b]{$~\epsilon_2$}
\Text(150,87)[bl]{$(b)$}
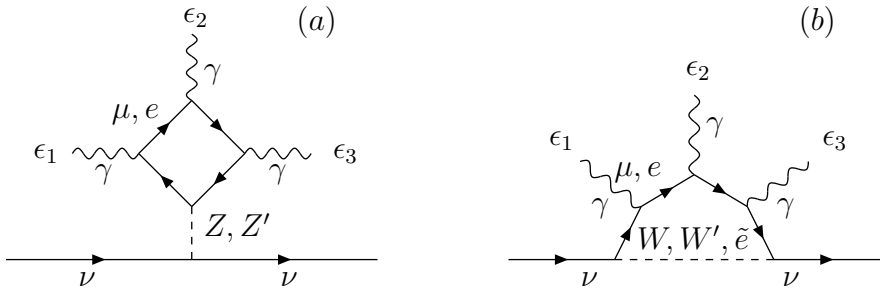
\efig{Fig. 1: Leading diagrams to five-leg 
photon--neutrino low-energy processes. 
The couplings and the exchanged particles 
depend on the  model.}

\end{document}